\documentclass[aps,prb,showpacs,onecolumn,groupedaddress]{revtex4-1}
\usepackage{times}
\usepackage{w-thm}
\usepackage{graphicx}
\usepackage[english]{babel}
\usepackage{blindtext}
\usepackage[utf8]{inputenc}
\usepackage{amsmath}
\usepackage{amssymb}
\usepackage[]{graphicx}
\begin{document}
\keywords{Interfaces, quantum kinetics, Boltzmann equation, boundary conditions}



\title[Boundary conditions]{About the quantum-kinetic derivation of boundary conditions for 
quasiparticle Boltzmann equations at interfaces}


\author{F. X. Bronold and F. Willert}
\affiliation{Institut f{\"ur} Physik,
             Universit{\"a}t Greifswald,
             17489 Greifswald,
             Germany}
\begin{abstract}
Quite a many electron transport problems in condensed matter physics are 
analyzed with a quasiparticle Boltzmann equation. For sufficiently slowly
varying weak external potentials it can be derived from the basic equations 
of quantum kinetics, provided quasiparticles can be defined and lead to a 
pole in the quantum-mechanical propagators. The derivation breaks down, 
however, in the vicinity of an interface which constitutes an abrupt strong
perturbation of the system. In this contribution we discuss in a tutorial 
manner a particular technique to systematically derive, for a planar, nonideal 
interface, matching conditions for the quasi-particle Boltzmann equation. 
The technique is based on pseudizing the transport problem by two auxiliary 
interface-free systems and matching Green functions at the interface. Provided 
quasiparticles exist in the auxiliary systems, the framework can be put onto 
the semiclassical level and the desired boundary conditions result. For ideal 
interfaces, the conditions can be guessed from flux conservation, but for 
complex interfaces this is no longer the case. The technique presented in 
this work is geared towards such interfaces.

\end{abstract}
\maketitle                   





\section{Introduction}

A cornerstone of nonequilibrium statistical mechanics is the Boltzmann equation.
Applying to situations where the motion of the constituents of a many-body system
can be split into free flights and randomly occurring scattering events, it 
provides the spatio-temporal evolution of the constituents' distribution 
function~\cite{Cercignani88}. Although rigorously valid only for classical dilute 
gases with short range interactions~\cite{Kogan69}, with the proper modifications and 
interpretations the Boltzmann equation can be also applied to classical ionized gases 
with long range interactions~\cite{Ecker72} as well as to the various microscopic 
degrees of freedom of condensed matter~\cite{SJ89}.

In the realm of transport processes in condensed matter, the Boltzmann equation
can be rigorously derived from quantum kinetics~\cite{Bogolyubov46,BG47,KB62,Keldysh65,Bonitz2016}
for electrons in metals at sufficiently low temperature where the electronic quasiparticles 
behave approximately like a weakly interacting dilute gas (see also the review 
by Rammer and Smith~\cite{RS86}). Originally performed by Landau~\cite{Landau58}, who 
also coined the name Fermi liquid for this type of electron system, the mapping 
of a degenerate interacting electron gas to a weakly interacting gas of quasiparticles 
can be systematically put in place within the quasiclassical theory of Fermi 
liquids~\cite{PK64,SR83}. The transport properties of normal as well as superfluid 
Fermi liquids turn out to be very well described by Boltzmann-type equations.  
For semiconductors and dielectrics the situation is different, since the transport
properties of interest are caused by nondegenerate conduction band electrons 
for which a quasiparticle description cannot be as rigorously derived as for 
metals. Still, under some assumptions, the Boltzmann equation applies~\cite{SL94,SL95},
justifying its wide use also for this class of materials~\cite{Roth92,SJ06}. In fact, 
it is the basis for the kinetic modeling of the miniaturized electronic semiconductor 
devices~\cite{FL88,RJH16} on which modern technology heavily relies.

Besides the presence of sufficiently long-lived quasiparticles, the second necessary
condition for the quantum kinetic derivation of an electron Boltzmann equation is a
sufficiently slowly varying external potential. In the vicinity of an interface, 
however, as it occurs at electronic contacts between different materials or at
the solid walls confining an ionized gas, this condition breaks down. Hence, 
additional considerations are necessary to obtain for interface electron transport 
problems a Boltzmann equation from the fundamental quantum kinetic description. 
Not surprisingly, it turns out that the Boltzmann equation has to be 
augmented by a boundary (or matching) condition at the interface. For superconducting 
metal contacts the boundary condition has been derived some time ago within the 
framework of the quasiclassical theory~\cite{Zaitsev84,Kieselmann85,MRS88} and ever 
since it has been applied in various forms~\cite{AKR86,Kieselmann87,CF01,GD07,ECB15}.
The boundary conditions for normal metal contacts arise as limiting cases, but 
they can be also obtained less rigorously from the single electron Schr\"odinger
equation and an heuristic construction of the distribution function via the 
single electron density matrix~\cite{Falkovsky83,DLP95}. For planar, ideal 
contacts, where the electron's lateral momentum is conserved, the matching 
conditions can be moreover intuitively guessed from flux conservation~\cite{Schroeder92}. 
In our work~\cite{BF17} concerning electron transport across a semiconductor-plasma
interface, as it occurs, for instance, in semiconductor-based microdischarges~\cite{EPC13},
we also employed this reasoning.  

The aim of the present work is to derive from quantum kinetics the boundary conditions
we used before and to extend them to nonideal interfaces. Having semiconductor-plasma 
interfaces in mind, we focus on nondegenerate electrons propagating across a planar 
interface containing surface roughness and/or surface defects. Instead of 
heuristically constructing the single electron density matrix, and deducing from 
it the matching conditions, we employ the method Kieselmann~\cite{Kieselmann85} 
developed for and applied with coworkers~\cite{AKR86,Kieselmann87} to superconducting 
metal contacts in thermal equilibrium. It is based on pseudizing the interface 
problem by two auxiliary interface-free systems and matching propagators at the 
interface with the help of surface Green functions~\cite{GR71}. In contrast to 
Kieselmann~\cite{Kieselmann85}, who pseudized the Eilenberger equation in Nambu 
space~\cite{Eilenberger68}, we pseudize the matrix Dyson equation for the Keldysh 
propagator required for the description of electron transport across the interface.
Within the constraints under which Boltzmann equations 
for the auxiliary systems can be derived quantum-kinetically, the pseudization 
approach can be put onto the semiclassical level resulting in boundary conditions 
for the halfspace's electron distribution functions. Although the conditions we 
end up with have been obtained before~\cite{DLP95}, the pseudization approach 
is more general. Being based on manybody Green functions, the renormalization of 
quasiparticles due to interaction processes may be systematically incorporated.
Besides its initial field of application, superconducting metal 
contacts~\cite{Zaitsev84,Kieselmann85,MRS88,Kieselmann87,AKR86,CF01,GD07,ECB15}, 
and the semiconductor-plasma interface we are looking at, the pseudization approach
works also for metallic~\cite{ZL98,AS16a,AS16b} as well as 
semiconducting~\cite{SK11} magnetic interfaces and may thus be also of interest 
to spintronic applications.

In the next section we present the derivation of the matching conditions using 
Kieselmann's~\cite{Kieselmann85} approach. From the conditions he obtains for 
superconducting proximity contacts, the conditions we are looking for cannot be 
read off. It is thus appropriate to provide a selfcontained derivation, although
some of the details can be also found in Kieselmann's thesis~\cite{Kieselmann85}. 
A concluding section puts our derivation into perspective and indicates routes 
of further studies.

\begin{figure}[t]
\includegraphics[width=\linewidth]{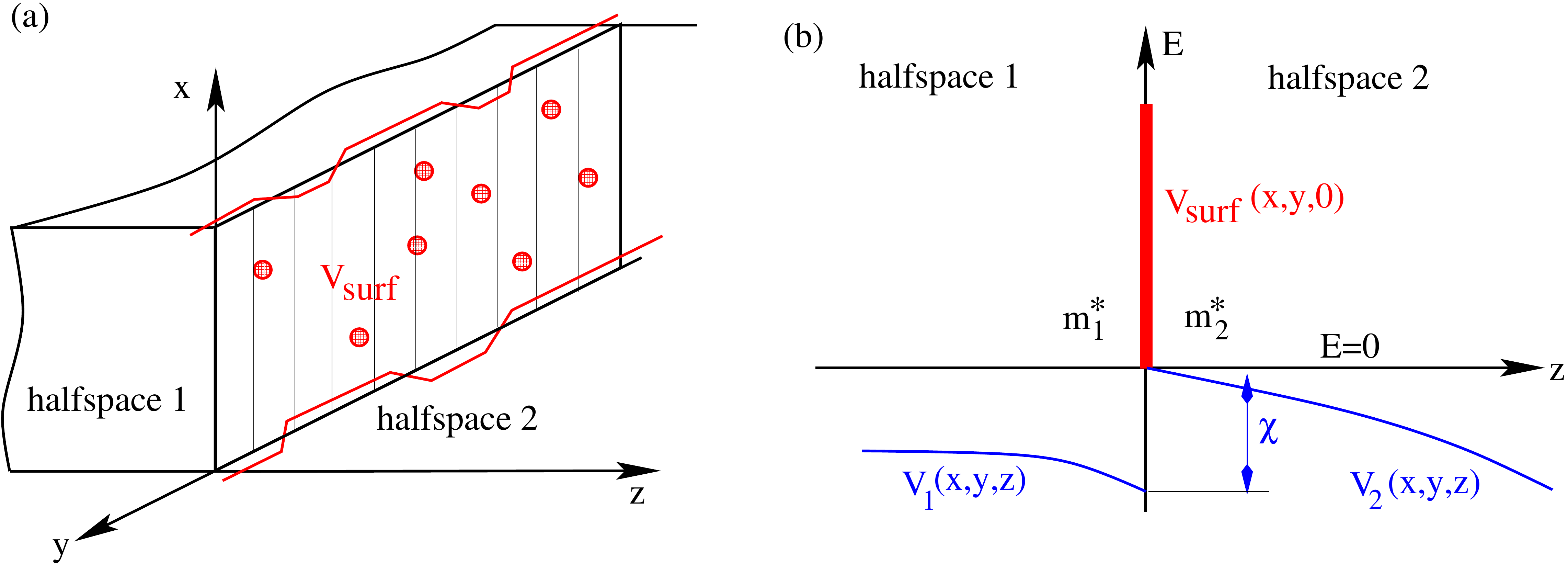}
\caption{\label{InterfaceCartoon}
(a) Illustration of a nonideal interface between a halfspace 1 and a halfspace 2.
Roughness and defects at the interface give rise to a random potential $V_{\rm surf}(x,y,0)$
located at $z=0$, which is assumed to be also the plane where the effective electron
mass $m^*(z)=m_1^*\Theta(-z)+m_2^*\Theta(z)$ and the external potential energy
$V_{\rm ext}(x,y,z)=V_1(x,y,z)\Theta(-z)+V_2(x,y,z)\Theta(z)$ change abruptly.
(b) The net profile of the potential energy seen by electrons crossing an unbiased
semiconductor-plasma interface. In addition to the surface potential $V_\mathrm{surf}$,
into which the nonideality of the interface is encoded, a potential step occurs  
at $z=0$ due to the electron affinity $\chi$ of the semiconductor. The variations 
away from $z=0$ are due to the electric field arising from charge separation across
the interface. The purpose of this work is to derive, starting from quantum kinetic equations,
conditions for matching electron distribution functions at $z=0$, as it is required 
for modeling interface electron transport by semiclassical Boltzmann equations.
}
\end{figure}

\section{Formalism}

In this section we derive the matching conditions for electron transport across 
the nonideal interface depicted in Fig.~\ref{InterfaceCartoon} using the approach 
developed by Kieselmann~\cite{Kieselmann85}. Gearing towards a description of 
semiconductor-metal or semiconductor-plasma interfaces, where electrons are 
typically nondegenerate, we cannot rely on the presence of a Fermi surface. The 
derivation of the matching conditions is thus subject to the same limitations 
as the derivation of the Boltzmann equation for nondegenerate electrons in 
semiconductors itself.  

\subsection{Dyson equation for the Keldysh matrix propagator}

We start with the Keldysh matrix propagator. In the notation of Rammer and 
Smith~\cite{RS86}, which we use throughout this work, it reads in terms 
of the retarded (R), advanced (A), and Keldysh (K) components,
\begin{align}
\underbar{G}(1,1^\prime) = \begin{pmatrix}
                                G^{\rm R}(1,1^\prime) & G^{\rm K}(1,1^\prime) \\
                                0 & G^{\rm A}(1,1^\prime) 
                        \end{pmatrix}
\label{Gatomic}
\end{align}
with $1=(\vec{r}_1, t_1)$ and $1^\prime=(\vec{r}^{\,\prime}_1, t^\prime_1)$. Defining  
the selfenergy
\begin{align}
\underline{\Sigma}(1,1^\prime) = \begin{pmatrix}
                                \Sigma^{\rm R}(1,1^\prime) & \Sigma^{\rm K}(1,1^\prime) \\
                                0 & \Sigma^{\rm A}(1,1^\prime)
                        \end{pmatrix}
\end{align}
and the $\otimes$ product, 
\begin{align}
(\underbar{A}\otimes\underbar{B})(1,1^\prime)=\int d2\, \underbar{A}(1,2)\,\underbar{B}(2,1^\prime)~,
\end{align}
the Dyson equation for the matrix propagator can be written in the form 
\begin{align}
[\underbar{G}_{\,0}^{-1} - \underline{\Sigma}]\otimes \underbar{G} = \underbar{E}~, 
\label{Dyson}
\end{align}
where $\underbar{E}=\delta(1-1^\prime)\,\underbar{1}$ and ($\hbar=1$)
\begin{align}
\underbar{G}_{\,0}^{-1}(1,1^\prime)=\bigg[i\frac{\partial}{\partial t_1}  + \frac{1}{2} \nabla_1 \frac{1}{m^*(1)}\nabla_1 
- V_{\rm ext}(1) - V_{\rm surf}(1)\bigg]\underbar{1}\,\delta(1-1^\prime)~
\label{G0}
\end{align}
with $m^*(1)=m_1^*\Theta(-z_1)+m_2^*\Theta(z_1)$, $V_{\rm ext}(1)=V_1(1)\Theta(-z_1)+V_2(1)\Theta(z_1)$, and
\begin{align}
V_{\rm surf}(1)=\int d^2 r_\parallel \bigg[ 
V_{\mathrm{s}_1}(\vec{r}_\parallel,t_1)\delta(\vec{r}_1-\vec{r}_{\mathrm{s}1})
+V_{\mathrm{s}_2}(\vec{r}_\parallel,t_1)\delta(\vec{r}_1-\vec{r}_{\mathrm{s}2}) \bigg]
\end{align}
with $\vec{r}_{\mathrm{s}_1}=(\vec{r}_\parallel,-\eta)$ and $\vec{r}_{\mathrm{s}_2}=(\vec{r}_\parallel,\eta)$,
where $\eta$ is infinitesimally small. Following Kieselmann~\cite{Kieselmann85}, the interface potentials 
$V_{\mathrm{s}_k}$ are infinitesimally off $z=0$, the geometrical position of the interface, defined by the 
abrupt changes of the material parameters (see Fig.~\ref{InterfaceCartoon}). The shifts are required for 
the construction of definite matching conditions for the distribution functions.  At the end of the 
construction  $\eta$ is set to zero leading to an interface potential characterized by a strength 
$V_{\mathrm{s}}=V_{\mathrm{s}_1}+V_{\mathrm{s}_2}$. 

Due to the continuity of the electron wave function, the propagator defined in~\eqref{Gatomic} 
is continuous across the interface if one stays on the atomic scale. The distribution function, 
however, which is an object applicable to the larger semiclassical scale, suffers a jump. It is 
our aim to calculate this jump, starting from the equations on the atomic scale given in this 
section.

\subsection{Pseudizing the interface transport problem}

\begin{figure}[t]
\includegraphics[width=\linewidth]{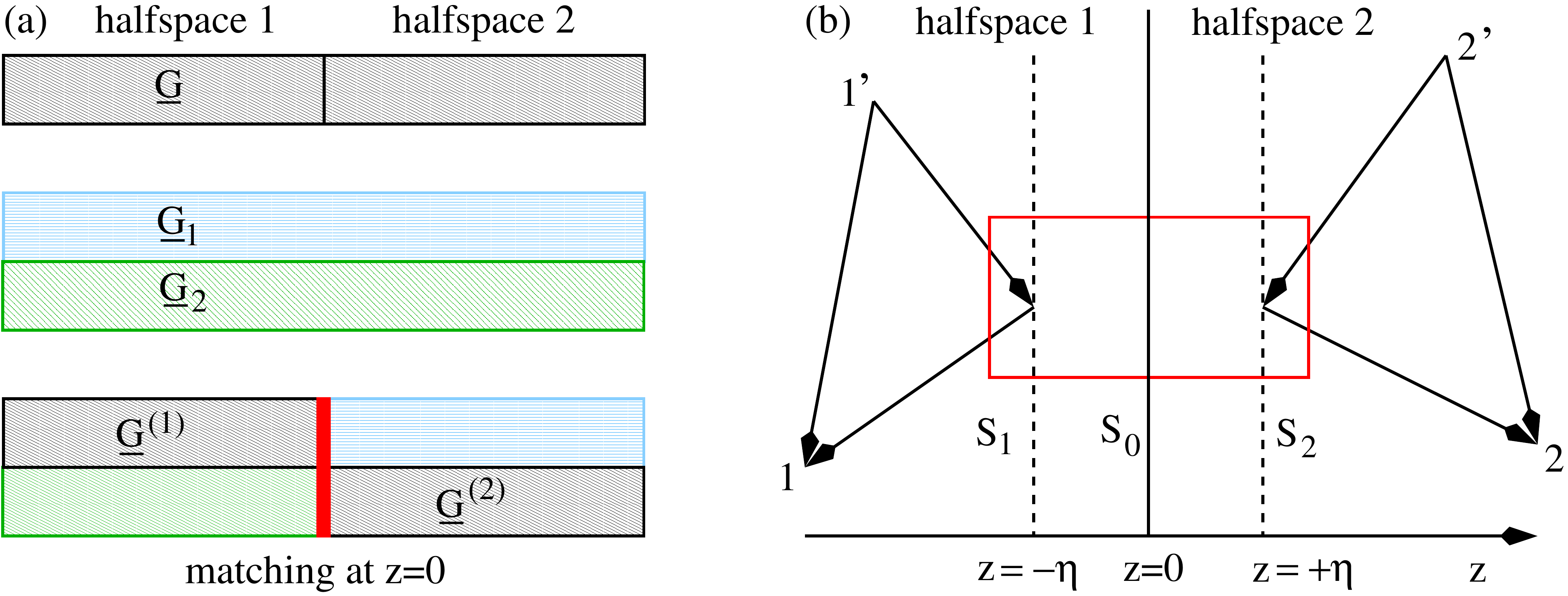}
\caption{\label{Pseudizing}
(a) Quintessence of the pseudization approach. The physical system, whose Green function
$\underbar{G}$ has to be determined, consists of two halfspaces, separated by an interface, 
and each with its own material parameters and potential energies, as shown in the first row. 
Instead of calculating $\underbar{G}$ directly, one considers initially two auxiliary, 
interface-free systems obtained by continuing the left and right side of the original system, 
respectively, to the other side of the interface, as depicted in the second row. With the 
auxiliary Green functions $\underbar{G}_1$ and $\underbar{G}_2$, a condition, shown in red 
in the third row, can be constructed, which connects the Green functions $\underbar{G}^{(1)}$ 
and $\underbar{G}^{(2)}$ applicable to the two sides of the original problem. Due to 
the flexibility of the continuation process, things can be arranged such that after matching
$\underbar{G}_k$ drops out and only $\underbar{G}^{(k)}$ remains ($k=1,2$). Put onto the 
semiclassical level, matching conditions for the electron distribution functions emerge 
from this approach. (b) The two possibilities within a halfspace an electron has to propagate 
between two points. Besides the direct path there is also a path involving a point at the 
interface. To construct matching conditions for the distribution functions, two auxiliary 
planes $S_1$ and $S_2$, shifted infinitesimally to the left and the right of $S_0$, the 
plane where the abrupt changes of the material parameters occur, are required. The matching
condition takes care of the processes occurring within the red box and in effect eliminates 
the direct propagation of an electron between the two halfspaces. 
}
\end{figure}

A Boltzmann equation for the electron distribution function can be derived rigorously 
from~\eqref{Dyson} for weakly interacting electrons 
in a slowly varying external field. The latter condition is obviously not fulfilled 
in the presence of an interface, where material properties and potentials change on an 
atomic scale. Hence, the standard tool of deriving a Boltzmann equation from quantum 
kinetics, the gradient expansion, cannot be applied. It is however possible to cast the 
problem in a manner enabling the gradient expansion despite the rapid variation of the 
system parameters in the vicinity of the interface. 

The idea we describe in this work is based on pseudizing the interface by defining two 
auxiliary interface-free systems. Originally, the method arose in the context of
electronic structure calculations~\cite{GR71} but Kieselmann~\cite{Kieselmann85,Kieselmann87}
and his coworkers~\cite{AKR86} applied it to the Eilenberger equation~\cite{Eilenberger68} 
for Matsubara Green functions in Nambu space to study proximity effects at superconducting 
metal-metal contacts. We adopt this approach to deduce from~\eqref{Dyson} the matching 
conditions for the electron distribution function at a planar nonideal interface. The 
quintessence of the approach is illustrated in Fig.~\ref{Pseudizing}.

To obtain the two auxiliary systems, we continue the halfspace systems throughout the 
whole space, as shown in Fig.~\ref{Pseudizing}a. Denoting the propagators 
for the two auxiliary systems by $\underbar{G}_1$ and $\underbar{G}_2$, satisfying 
\begin{align}
[\underbar{G}_{0,k}^{-1} - \underline{\Sigma}_k]\otimes \underbar{G}_k = \underbar{E}~,
\label{Dyson_k}
\end{align}
where $\underline{\Sigma}_k$ is the selfenergy of the interface-free system $k=1,2$ and 
\begin{align}
G_{0,k}^{-1}(1,1^\prime)=\bigg[i\frac{\partial}{\partial t_1}  + \frac{1}{2m_k^*} \Delta
- V_k(1)\bigg]\,\underbar{1}\,\delta(1-1^\prime)~,
\label{G0k}
\end{align}
we can make the Ansatz
\begin{align}
\underbar{G}=\underline{G}^{(1)}\Theta(-z)+\underline{G}^{(2)}\Theta(z) 
\end{align}
with 
\begin{align}
\underbar{G}^{(k)}=\underbar{G}_k + \underbar{G}_k\otimes \underbar{T}_k \otimes \underbar{G}_k~
\label{DefTmatrix}
\end{align}
and an interface T-matrix $\underbar{T}_k$ to be determined from the atomic scale
Dyson equations~\eqref{Dyson} and~\eqref{Dyson_k}. The two terms on the right hand side of the last 
equation reflect the two possibilities, shown in Fig.~\ref{Pseudizing}b,
an electron has within a halfspace to reach two of its points.

Since the auxiliary propagators $\underbar{G}_k$ apply to interface-free systems, they 
can be connected to electron distribution functions in the standard manner and under the 
standard restrictions concerning the existence of quasiparticles and the spatio-temporal 
variation of the potentials~\cite{KB62}. The abrupt changes are confined to the 
interface T-matrix. Following standard procedure in the notation of Rammer and
Smith~\cite{RS86}, we hence apply the left-right trick, that is, we 
multiply~\eqref{DefTmatrix} from the left with $\underbar{G}_k^{-1}$ and subtract from 
it~\eqref{DefTmatrix} after it was multiplied from the right with $\underbar{G}_k^{-1}$. 
With~\eqref{Dyson_k}, the result is 
\begin{align}
[\underbar{G}_{k,0}^{-1} - \underline{\Sigma}_k, \underline{G}^{(k)}]_\otimes = 
[\underbar{T}_k, \underbar{G}_k]_\otimes ~.
\label{LeftRight}
\end{align}
Utilizing the slow spatio-temporal variation of $\underbar{G}_k$ and $\underline{\Sigma}_k$, 
a first-order gradient expansion~\cite{RS86} applies to~\eqref{LeftRight} yielding for the 
Keldysh component,
\begin{align}
[G_{k,0}^{-1} &- \mathrm{Re}\Sigma_k, G^{(k)K}]_p - [\Sigma_k^K, \mathrm{Re}G^{(k)}]_p 
- \Sigma_k^K A^{(k)} +\Gamma_k G^{(k)K} = \frac{1}{2}[T_k^R + T_k^A,G_k^K]_p 
\nonumber\\
&+ \frac{1}{2}[T_k^K, G_k^R+G_K^A]_p
+ T_k^K A_k - i(T_k^R-T_k^A)G_k^K~,
\label{KeldyshComponent}
\end{align}
and for the difference of the diagonal components,
\begin{align}
[G_{k,0}^{-1}- \mathrm{Re}\Sigma_k,A^{(k)}]_p - [\Gamma_k,\mathrm{Re}G^{(k)}]_p =
\frac{1}{2}[T_k^R + T_k^A,A_k]_p
-i[T_k^R - T_k^A,\mathrm{Re}G_k]_p~,
\label{RetAdComponent}
\end{align}
where we introduced, in accordance with standard procedure~\cite{RS86}, the spectral function,
\begin{align}
A^{(k)} &= i(G^{(k)R}-G^{(k)A})~,
\label{Ak}
\end{align}
and the combinations
\begin{align}
\Gamma_k &= i(\Sigma_k^R - \Sigma_k^A)~~,~~
\mathrm{Re}\Sigma_k = \frac{1}{2}(\Sigma_k^R + \Sigma_k^A)~~,~~ \\
\mathrm{Re}G^{(k)} &= \frac{1}{2}(G^{(k)R}+G^{(k)A)})~~,~~ 
\mathrm{Re}G_k = \frac{1}{2}(G_k^R+G_k^A)~.
\end{align}

The functions entering~\eqref{KeldyshComponent} and~\eqref{RetAdComponent} are the 
Fourier transforms of the original functions with respect to the difference variables,
$t=t_1-t_1^\prime$ and $\vec{r}=\vec{r}_1-\vec{r}_1^{\,\prime}$, that is, they are all
of the form 
\begin{align}
F(\vec{R},T,\vec{p},E) = \int d^3 r \int dt \exp[iEt - i\vec{p}\cdot\vec{r}\,] 
F(\vec{R}+\vec{r}/2,T+t/2,\vec{R}-\vec{r}/2,T-t/2)~,
\label{QCtransform}
\end{align}
where $\vec{R}=(\vec{r}_1+\vec{r}_1^{\,\prime})/2$ and $T=(t_1+t_1^\prime)/2$ are sum variables.
The Poisson bracket reads 
\begin{align}
[A,B]_p = \frac{\partial A}{\partial E}\frac{\partial B}{\partial T} 
        - \frac{\partial A}{\partial T}\frac{\partial B}{\partial E}
        + (\nabla_R A)\cdot(\nabla_p B) 
        - (\nabla_p A)\cdot(\nabla_R B)~,
\end{align}
where the subscripts of the Nabla operator indicates whether it affects the vector $\vec{R}$ or 
the vector $\vec{p}$. 

Since our goal is to derive the matching conditions for the distribution functions, which do 
not contain derivatives, at least in the form arising from the single electron Schr\"odinger 
equation~\cite{Falkovsky83}, we drop the Poisson brackets involving the interface T-matrix. 
Equation~\eqref{RetAdComponent} for the spectral function has then the canonical form studied 
by Kadanoff and Baym~\cite{KB62}. Its solution is given by 
\begin{align}
A^{(k)}(\vec{R},T,\vec{p},E) &= 
\frac{\Gamma_k(\vec{R},T,\vec{p},E)}
{[E-\frac{\vec{p}^{\,2}}{2m_k^*}-V_k(\vec{R},T,\vec{p},E)-\mathrm{Re}\Sigma_k(\vec{R},T,\vec{p},E)]^2
+[\frac{\Gamma_k(\vec{R},T,\vec{p},E)}{2}]^2}
\nonumber\\
&=A_k(\vec{R},T,\vec{p},E)~.
\label{SpectralFunction}
\end{align}
Hence, only the spectral functions of the auxiliary systems are required. The equation for 
the Keldysh component, which eventually becomes the Boltzmann equation, reads, after ignoring
Poisson brackets containing the interface T-matrix,
\begin{align}
[G_{k,0}^{-1} &- \mathrm{Re}\Sigma_k, G^{(k)K}]_p - [\Sigma_k^K, \mathrm{Re}G^{(k)}]_p
- \Sigma_k^K A^{(k)} +\Gamma_k G^{(k)K} = 
T_k^K A_k - i(T_k^R-T_k^A)G_k^K~.
\label{Pre_Boltzmann}
\end{align}

Up to this point it was not necessary to specify the properties of the interface-free 
systems. However, to proceed we now have to assume that the propagators 
$\underbar{G}_k$ have a quasiparticle pole. Only then is it possible to derive 
from~\eqref{Pre_Boltzmann} a Boltzmann-type kinetic equation for the electron 
distribution function. The range of validity of the distribution function's 
matching conditions is thus the one of the Boltzmann equation itself.

For metals, for instance, where quasiparticles are well defined in the vicinity of the 
Fermi surface, the derivation of the matching condition can be pushed through rather 
rigorously as long as the thermal energy and the excitation energies are much smaller  
than the Fermi energy and the external potential varies on a scale much larger than  
the inverse of the Fermi wave number. The result is a matching condition for the 
$\xi-$integrated propagator. We are however interested in interfaces where at least 
one side is semiconducting. The derivation of the matching conditions can thus not 
be based on the presence of a Fermi scale. Instead it rests on the same ad hoc 
approximations as the derivation of the Boltzmann equation for a homogeneous semiconductor.
As for the Boltzmann equation itself, which is successfully used for the modeling of 
electron transport in semiconductors, although it cannot be rigorously justified from a
quantum kinetic point of view, the matching conditions we obtain may have a wider range 
of validity than the derivation suggests.

To avoid a discussion of the selfenergies $\underline{\Sigma}_k$, which affects the 
analytical structure of the matrix propagator, we proceed by considering the simplest 
situation: a weakly interacting electron gas in a sufficiently slowly varying external
field (besides the discontinuities at the interface). Hence, the spectral 
function~\eqref{SpectralFunction} becomes 
\begin{align}
A_k(\vec{R},T,\vec{p},E)=2\pi\delta\bigg(E-\frac{\vec{p}^{\,\,2}}{2m_k^*}-V_k(\vec{R},T)\bigg)~,
\label{Ak0}
\end{align}
which together with 
\begin{align}
G^{(k)K}(\vec{R},T,\vec{p},E)=-iA_k(\vec{R},T,\vec{p},E)h^{(k)}(\vec{R},T,\vec{p}) 
\label{GKk}
\end{align}
leads to
\begin{align}
h^{(k)}(\vec{R},T,\vec{p})=i\int\frac{dE}{2\pi} G^{(k)K}(\vec{R},T,\vec{p},E)~
\end{align}
and a similar equation for $G_k^K$. Retarded and advanced Green functions, 
$G_k^R$ and $G_k^A$, required for the calculation of the interface T-matrix, follow straight 
from~\eqref{Ak} using~\eqref{Ak0}, resulting in 
\begin{align}
G_k^{R}(\vec{R},T,\vec{p},E)=[G_k^{A}(\vec{R},T,\vec{p},E)]^*=
\bigg(E+i\eta - \frac{\vec{p}^{\,2}}{2m_k^*} -V_k(\vec{R},T)\bigg)^{-1}~.
\label{GRk}
\end{align}

To be consistent with the spectral function~\eqref{Ak0}, the Poisson brackets on the 
left hand side of~\eqref{Pre_Boltzmann}, affecting $\mathrm{Re}G^{(k)}$ and $\mathrm{Re}\Sigma_k$, 
will be also ignored. Applying then $i\int dE/(2\pi)$ on both sides of the emerging 
equation yields
\begin{align}
\bigg[\frac{\partial}{\partial T}+\frac{\vec{p}}{m_k^*}\cdot\nabla_R
-\nabla_R V_k(\vec{R},T)\cdot \nabla_p \bigg]h^{(k)}(\vec{R},T,\vec{p})
-C_k(\vec{R},T,\vec{p}) 
=\delta(Z)U_k(\vec{R},T,\vec{p})~,
\label{BE1}
\end{align}
with the collision integral 
\begin{align}
C_k(\vec{R},T,\vec{p}) = i\Sigma_k^K(\vec{R},T,\vec{p},\tilde{E}_k)
                       - \Gamma_k(\vec{R},T,\vec{p},\tilde{E}_k) h^{(k)}(\vec{R},T,\vec{p})
\label{CollIntegral}
\end{align}
and the inhomogeneity on the right hand side 
\begin{align}
U_k(\vec{R},T,\vec{p})=i\tau_{\,k}^K(\vec{R}_\parallel,T,\vec{p}_\parallel,\tilde{E}_k)
-i[\tau_{\,k}^R(\vec{R}_\parallel,T,\vec{p}_\parallel,\tilde{E}_k) 
- \tau_{\,k}^A(\vec{R}_\parallel,T,\vec{p}_\parallel,\tilde{E}_k)]h_k(\vec{R},T,\vec{p})~,
\label{Inhomogeneity}
\end{align}
where $\tilde{E}_k=\tilde{E}_k(\vec{R},T,\vec{p})=\vec{p}^{\,2}/2m_k^*+V_k(\vec{R},T)$,
$\vec{R}_\parallel=(X,Y)$, and $\vec{p}_\parallel=(p_x,p_y)$. Using~\eqref{Ak0} also
for the internal propagators of the selfenergies making up the collision integral
leads to an equation for the function $h^{(k)}$ in terms of the function $h_k$. 

The components of the function $\underline{\tau}_{\,k}$, making up the inhomogeneity $U_k$ of
the Boltzmann equation~\eqref{BE1}, are the  projections of the components of the 
function $\underbar{T}_k$ onto the interface plane $S_k$. Using summation convention, we write 
\begin{align}
\underbar{T}_k(\vec{r},t,\vec{r}^{\,\prime}t^\prime)=\delta(\vec{r}-\vec{r}_{\mathrm{s}_k})
\delta(t-t_{\mathrm{s}_k})
\underline{\tau}_{\,k}(\vec{r}_{\mathrm{s}_k},t_{\mathrm{s}_k},\vec{r}_{\mathrm{s}_k}^{\,\prime},t_{\mathrm{s}_k}^\prime)
\delta(\vec{r}_{\mathrm{s}_k}^{\,\prime}-\vec{r}^{\,\prime})
\delta(t_{\mathrm{s}_k}^\prime-t^\prime)~,
\end{align}
which yields by applying the transformation~\eqref{QCtransform}
\begin{align}
\underbar{T}_k(\vec{R},T,\vec{p},E)=\delta(Z)\underline{\tau}_{\,k}(\vec{R}_\parallel,T,\vec{p}_\parallel,E)
\end{align}
with
\begin{align}
\underline{\tau}_{\,k}(\vec{R}_\parallel,T,\vec{p}_\parallel,E)=
\int\!\! d^2 r_\parallel \int \!dt \exp[iEt-i\vec{p}_\parallel\cdot\vec{r}_\parallel]
\underline{\tau}_{\,k}(\vec{R}_\parallel+\vec{r}_\parallel/2,T+t/2,\vec{R}_\parallel-\vec{r}_\parallel/2,T-t/2)~.
\end{align}
Thus, the matching conditions for the functions $h^{(k)}$, and hence 
for the distribution functions $f^{(k)}$, which are closely related to it, as we will see below, are hidden 
in the inhomogeneity $U_k$, which in turn depends on the interface T-matrix $\underline{\tau}_{\,k}$. The latter  
is the central object of the pseudization approach and will be calculated in the next section.

\subsection{Calculation of the function $\underline{\tau}_{\,k}(\vec{R}_\parallel,T,\vec{p}_\parallel,E)$}

We now adopt Kieselmann's strategy~\cite{Kieselmann85} to calculate the surface projection of the 
interface T-matrix to the situation we are interested in. Initially, the calculation stays on the 
atomic scale, at the end however, the result will be put onto the semiclassical scale by a 
first order gradient expansion.

Projecting~\eqref{DefTmatrix} onto the plane $S_k$ yields, after multiplying from the left and the right 
with the inverse $\underline{{\cal G}}_k^{-1}$ of the surface projection of the propagator of the 
interface-free system,
\begin{align}
\underline{{\cal G}}_k(\vec{r}_\parallel,t,\vec{r}_\parallel^{\,\prime},t^\prime)=
\underbar{G}_k(\vec{r}_{\mathrm{s}_k},t,\vec{r}_{\mathrm{s}_k}^{\,\prime},t^\prime)~,
\end{align}
the useful equation 
\begin{align}
\underline{\tau}_{\,k}(\vec{r}_\parallel,t,\vec{r}_\parallel^{\,\prime},t^\prime)&=
\underline{{\cal G}}_k^{-1}(\vec{r}_\parallel,t,\vec{r}_\parallel^{\,\prime\prime},t^{\prime\prime})
\nonumber\\
&\circ\bigg[\underline{{\cal G}}^{(k)}(\vec{r}_\parallel^{\,\prime\prime},
t^{\prime\prime},\vec{r}_\parallel^{\,\prime\prime\prime},t^{\prime\prime\prime})
-\underline{{\cal G}}_k(\vec{r}_\parallel^{\,\prime\prime},t^{\prime\prime},
\vec{r}_\parallel^{\,\prime\prime\prime},t^{\prime\prime\prime})\bigg]
\circ\underline{{\cal G}}_k^{-1}(\vec{r}_\parallel^{\,\prime\prime\prime},t^{\prime\prime\prime},
\vec{r}_\parallel^{\,\prime},t^\prime)~,
\label{ProjectionTmatrix}
\end{align}
where $\circ$ is the $\otimes$ product for the lateral spatial coordinates and times only. 
The interface T-matrix can thus be determined from the surface projections of the propagators of the 
two interface-free auxiliary systems, provided an expression for the surface projection
\begin{align}
\underline{{\cal G}}^{(k)}(\vec{r}_\parallel,t,\vec{r}_\parallel^{\,\prime},t^\prime)= 
\underbar{G}^{(k)}(\vec{r}_{\mathrm{s}_k},t,\vec{r}_{\mathrm{s}_k}^{\,\prime},t^\prime)
\label{Gprojection}
\end{align}
can be also worked out. 

As for the selfenergy $\underline{\Sigma}_k$, we compute $\underline{{\cal G}}^{(k)}$ and thus
the interface T-matrix $\underline{\tau}_{\,k}$ with the spectral function~\eqref{Ak0}. Hence, the 
starting point of the calculation are the Dyson 
equations~\eqref{Dyson} and~\eqref{Dyson_k} without the selfenergies. Writing
Eq.~\eqref{Dyson_k} in the conjugated form, the two equations we start from read
\begin{align}
\underbar{G}_{\,0}^{-1}\otimes\underbar{G}^{(k)} &=\underbar{E}~,
\label{Eq1}\\
\underbar{G}_k\otimes\overleftarrow{\underbar{G}}_{0,k}^{-1} &=\underbar{E}~,
\label{Eq2}
\end{align}
where the overleftarrow in the second equation indicates that the differential operators inside 
$\underbar{G}_{0,k}^{-1}$ act to the left. Multiplying now Eq.~\eqref{Eq1} from the left with
$\underbar{G}_k$ and subtracting from it Eq.~\eqref{Eq2}, premultiplied from the right with 
$\underbar{G}^{(k)}$, yields after integrations over the halfspace $R_k$ and the time axis 
the following equation~\cite{Kieselmann85}
\begin{align}
\underbar{G}_k(\vec{r}^{\,\prime},t^\prime,\vec{r}^{\,\prime\prime},t^{\prime\prime})
&-\underbar{G}^{(k)}(\vec{r}^{\,\prime},t^\prime,\vec{r}^{\,\prime\prime},t^{\prime\prime})=\int_{R_k} d^3 r \int dt\,
\underbar{G}_k(\vec{r}^{\,\prime},t^\prime,\vec{r},t)
\nonumber\\
&\times\bigg[\frac{1}{2}\nabla_r\frac{1}{m_k^*}\nabla_r-V_{\rm surf}(\vec{r},t)-\frac{1}{2m_k^*}\overleftarrow{\Delta}_r\bigg]
\underbar{G}^{(k)}(\vec{r},t,\vec{r}^{\,\prime\prime},t^{\prime\prime})~.
\end{align}
The term involving the time derivative dropped out after a partial integration because the 
time dependence of the components of the matrix propagator eliminates the boundary term of
the time integration. 

Setting, following Kieselmann~\cite{Kieselmann85}, $\vec{r}^{\,\prime}=\vec{r}_{\mathrm{s}_k}^{\,\prime}$ 
and $\vec{r}^{\,\prime\prime}=\vec{r}_{\mathrm{s}_k}^{\,\prime\prime}$, produces, after applying 
the second Green theorem, an equation connecting the surface projections
$\underline{{\cal G}}_k$ of the interface-free systems with the surface projections
$\underline{{\cal G}}^{(k)}$ of the interface-carrying system. Defining 
\begin{align}
\underline{{\cal E}}(1,1^\prime)&=\delta(\vec{r}_\parallel-\vec{r}_\parallel^{\,\prime})\delta(t-t^\prime)\,\underbar{1}~,
\\
^\prime\underline{{\cal G}}^{(k)} &=\frac{1}{2m_k^*}\frac{\partial}{\partial z}
\underbar{G}^{(k)}(\vec{r}_{\mathrm{s}_k},t,\vec{r}_{\mathrm{s}_k}^{\,\prime\prime},t^{\prime\prime})~,
\\
\underline{{\cal G}}_k^{\,\prime} &= \frac{1}{2m_k^*}\frac{\partial}{\partial z} 
\underbar{G}_k(\vec{r}_{\mathrm{s}_k}^{\,\prime\prime}~,
t^{\prime\prime},\vec{r}_{\mathrm{s}_k},t)~,
\end{align}
where in the last two definitions the position of the prime indicates whether the derivative is taken 
with respect to the first or second spatial variable, the desired relation reads
\begin{align}
	\bigg[\underline{{\cal G}}^{-1}_k \circ \bigg(\underline{{\cal E}} + (-1)^{k} \underline{{\cal G}}_k^{\,\prime}\bigg) 
-  V_{\mathrm{s}_k} \bigg]\circ \underline{{\cal G}}^{(k)}=  
\underline{{\cal E}}\, +\,
^\prime\underline{{\cal G}}^{(k)}(-1)^{k}~.
\label{Important}
\end{align} 
Performing finally the limit $\eta\rightarrow 0$ and using the matching conditions for the surface projections 
onto $S_0$, 
\begin{align}
^{(+)}\underline{{\cal G}}_k^{\,\prime} \,-\, ^{(-)}\underline{{\cal G}}_k^{\,\prime} &= -\underline{\cal{E}}~,
\label{M1}\\
^\prime\underline{{\cal G}}^{(2)(+)} \,-\, ^{\prime}\underline{{\cal G}}^{(1)(-)} &= -\underline{\cal{E}}~,
\label{M2}\\
\underline{{\cal G}}^{(1)}=\underline{{\cal G}}^{(2)} &= \underline{{\cal G}}~,
\label{M3}
\end{align}
where the superscripts $^{(-)}$ and $^{(+)}$ indicate whether in the limit $\eta\rightarrow 0$ the plane 
$S_0$ is approached from the left or the right, and summing up~\eqref{Important} for $k=1$ and $k=2$ yields
\begin{align}
\bigg[\underline{{\cal G}}_2^{-1}\circ\, ^{(-)}\underline{{\cal G}}_2^{\,\prime}
\,-\,\underline{{\cal G}}_1^{-1}\circ\, ^{(+)}\underline{{\cal G}}_1^{\,\prime}
-V_s\bigg]
\circ\underline{{\cal G}}= \underline{{\cal E}}~,
\label{SurfaceProjectionG}
\end{align}
where we set $V_s=V_{\mathrm{s}_1}+V_{\mathrm{s}_2}$. Since all functions appearing in the bracket on the
left hand side of the above equation are known, inverting the bracket yields the surface projection 
$\underline{{\cal G}}$ required for the calculation of $\underline{\tau}_{\,k}$. 

The matching conditions~\eqref{M1}--\eqref{M3} can be obtained from the interaction-free Dyson
equations for $\underbar{{\cal G}}_k$ and $\underbar{{\cal G}}$ following the strategy developed 
by Kieselmann~\cite{Kieselmann85}. Since the conditions are rather important we give the details.
In order to derive~\eqref{M2} one integrates $\underbar{G}_{\,0}^{-1}\otimes\underbar{G}=\underbar{E}$ 
along $z$ from $S_1$ to $S_0$ choosing, respectively, the starting and end point of the z-integration 
infinitesimally smaller than $z_{\mathrm{s}_1}$ and $z_{\mathrm{s}_0}$ while setting the second spatial 
variable to $\vec{r}_{\mathrm{s}_k}^{\,\prime}$. The result is (suppressing the time variables)
\begin{align}
\frac{1}{2m_1^*}\frac{\partial}{\partial z}\,
\underline{{\cal G}}(\vec{r}_{\mathrm{s}_0}^{\,\,0^-},\vec{r}_{\mathrm{s}_k}^{\,\prime})
-\frac{1}{2m_1^*}\frac{\partial}{\partial z}\,
\underline{{\cal G}}(\vec{r}_{\mathrm{s}_1}^{\,\,0^-},\vec{r}_{\mathrm{s}_k}^{\,\prime})
=\delta(\vec{r}_\parallel-\vec{r}_\parallel^{\,\prime})\delta(t-t^\prime)\underbar{1}\delta_{1k}
+V_{\mathrm{s}_1}\underline{{\cal G}}(\vec{r}_{\mathrm{s}_1},\vec{r}_{\mathrm{s}_k}^{\,\prime})~,
\label{Connect1}
\end{align}
where the superscript of the vector $\vec{r}_{\mathrm{s}_k}$ indicates the 
infinitesimal shift of the $z-$component of the vector with respect to the position 
of the plane $S_k$. Integrating, on the other hand, the Dyson equation across $z=0$ without 
crossing $S_1$ and $S_2$ gives after projection onto $S_0$ 
\begin{align}
\frac{1}{2m_1^*}\frac{\partial}{\partial z}\,
\underline{{\cal G}}(\vec{r}_{\mathrm{s}_0}^{\,\,0^-},\vec{r}_{\mathrm{s}_k}^{\,\prime})
=\frac{1}{2m_2^*}\frac{\partial}{\partial z}\,
\underline{{\cal G}}(\vec{r}_{\mathrm{s}_0}^{\,\,0^+},\vec{r}_{\mathrm{s}_k}^{\,\prime})
\label{Connect2}
\end{align}
since the surface potentials are not embraced by the procedure.  
Inserting~\eqref{Connect2} into~\eqref{Connect1} for $k=2$ yields 
\begin{align}
\frac{1}{2m_2^*}\frac{\partial}{\partial z}\,
\underline{{\cal G}}(\vec{r}_{\mathrm{s}_0}^{\,\,0^+},\vec{r}_{\mathrm{s}_2}^{\,\prime})
-\frac{1}{2m_1^*}\frac{\partial}{\partial z}\,
\underline{{\cal G}}(\vec{r}_{\mathrm{s}_1}^{\,\,0^-},\vec{r}_{\mathrm{s}_2}^{\,\prime})
=V_{\mathrm{s}_1}\underline{{\cal G}}(\vec{r}_{\mathrm{s}_1},\vec{r}_{\mathrm{s}_2}^{\,\prime})~,
\end{align}
which combined with~\eqref{Connect1} for $k=1$ leads to 
\begin{align}
\frac{1}{2m_2^*}\frac{\partial}{\partial z}\,
\underline{{\cal G}}(\vec{r}_{\mathrm{s}_0}^{\,\,0^+},\vec{r}_{\mathrm{s}_2}^{\,\prime})
-\frac{1}{2m_1^*}\frac{\partial}{\partial z}\,
\underline{{\cal G}}(\vec{r}_{\mathrm{s}_0}^{\,\,0^-},\vec{r}_{\mathrm{s}_1}^{\,\prime})
&= -\delta(\vec{r}_\parallel-\vec{r}_\parallel^{\,\prime})\delta(t-t^\prime)\underbar{1} \nonumber\\
&+ V_{\mathrm{s}_1}\underline{{\cal G}}(\vec{r}_{\mathrm{s}_1},\vec{r}_{\mathrm{s}_2}^{\,\prime})
-V_{\mathrm{s}_1}\underline{{\cal G}}(\vec{r}_{\mathrm{s}_1},\vec{r}_{\mathrm{s}_1}^{\,\prime}) 
\nonumber\\
&+\frac{1}{2m_1^*}\frac{\partial}{\partial z}\,
\underline{{\cal G}}(\vec{r}_{\mathrm{s}_1}^{\,\,0^-},\vec{r}_{\mathrm{s}_2}^{\,\prime})
-\frac{1}{2m_1^*}\frac{\partial}{\partial z}\,
\underline{{\cal G}}(\vec{r}_{\mathrm{s}_1}^{\,\,0^-},\vec{r}_{\mathrm{s}_1}^{\,\prime})~.
\end{align}
In the limit $\eta\rightarrow 0$, $S_k \rightarrow S_0$ and due to 
continuity of the Green functions on the atomic scale, which is also condition~\eqref{M3},
we obtain~\eqref{M2}. Finally, condition~\eqref{M1} follows for $k=1$ by applying the same 
procedure to $\underbar{G}_{0,1}^{-1}\otimes\underbar{G}_1=\underbar{E}$ while for $k=2$ it is 
obtained by integrating $\underbar{G}_{0,2}^{-1}\otimes\underbar{G}_2=\underbar{E}$ from $S_2$ 
towards $S_0$. 

Going back to~\eqref{SurfaceProjectionG}, we now solve this equation in first order gradient 
expansion. The $\circ$ operation is then simply a matrix product for functions of the form 
\begin{align}
\underline{{\cal F}}(\vec{R}_\parallel,T,\vec{p}_\parallel,E)=
\underbar{F}(\vec{R},T,\vec{p}_\parallel,z,E)~,
\label{Trafo1}
\end{align}
where on the right hand side $\vec{R}=(\vec{R}_\parallel,0)$, $z=0$, and 
\begin{align}
\underbar{F}(\vec{R},T,\vec{p}_\parallel,z,E)
=\int \frac{dp_z}{2\pi}\exp[-ip_z z]\underbar{F}(\vec{R},T,\vec{p},E)~.
\label{Trafo2}
\end{align}
To be able to perform this transformation is crucial for the construction of the matching 
conditions. It eventually connects the surface projections with the functions $h_k$. The 
pole structure of the propagators in turn is essential for this step. Without a pole, the 
backtransformation~\eqref{Trafo2} cannot be performed. Since we work with~\eqref{Ak0}, the 
transformation is straightforward. Treating a somewhat more general case, based on the 
extended quasiparticle approximation~\cite{SL94,SL95}, may be also possible, but then 
the interaction selfenergies have to be calculated and analyzed explicitly. This case 
will thus not be pursued in the present work.

For the spectral function~\eqref{Ak0} the components of the interaction-free propagator 
$\underbar{G}_k$ are given by~\eqref{GKk} and~\eqref{GRk}. Transforming the expressions
according to~\eqref{Trafo1} and~\eqref{Trafo2}, we obtain for the surface projections
of the interaction-free propagators and their derivatives, 
\begin{align}
\underline{{\cal G}}_k(\vec{R}_\parallel,T,\vec{p}_\parallel,E)&=\frac{1}{iv_{kz}}\begin{pmatrix}
                                1 & s_k(\vec{R}_\parallel,T,\vec{p}_\parallel,E)\\
                                0 & -1
                        \end{pmatrix}~, \label{Gk}\\
^{(\pm)}\underline{{\cal G}}^{\,\prime}_k(\vec{R}_\parallel,T,\vec{p}_\parallel,E)&=-\frac{1}{2}\begin{pmatrix}
                          \pm1 & d_k(\vec{R}_\parallel,T,\vec{p}_\parallel,E)\\
                                0 & \pm 1
                        \end{pmatrix} 
\label{Gkp}
\end{align}
with 
\begin{align}
s_k(\vec{R}_\parallel,T,\vec{p}_\parallel,E) &= h_k^+(\vec{R}_\parallel,T,\vec{p}_\parallel,E) 
+ h^-_k(\vec{R}_\parallel,T,\vec{p}_\parallel,E)~,
\label{sk}\\
d_k(\vec{R}_\parallel,T,\vec{p}_\parallel,E) &=h^+_k(\vec{R}_\parallel,T,\vec{p}_\parallel,E)
- h^-_k(\vec{R}_\parallel,T,\vec{p}_\parallel,E)~,
\label{dk}
\end{align}
where 
\begin{align}
h^+_k(\vec{R}_\parallel,T,\vec{p}_\parallel,E) &= 
h_k(\vec{R}_\parallel,T,\vec{p}_\parallel+|p_{kz}|\hat{e}_z)~,
\\
h^-_k(\vec{R}_\parallel,T,\vec{p}_\parallel,E) &=
h_k(\vec{R}_\parallel,T,\vec{p}_\parallel-|p_{kz}|\hat{e}_z)~,
\end{align}
and 
\begin{align}
|p_{kz}|(\vec{R}_\parallel,T,\vec{p}_\parallel,E) = m_k^* v_{kz}(\vec{R}_\parallel,T,\vec{p}_\parallel,E)
=\sqrt{2m_k^*[E-V_k(\vec{R}_\parallel,T)]-{\vec{p}_\parallel}^{\,2}}~.
\end{align}
Inserting~\eqref{Gk} and~\eqref{Gkp} into~\eqref{SurfaceProjectionG} we thus get
\begin{align}
\underline{\cal G}^{-1}=\underline{\cal G}_{\,0}^{-1} - V_s \, \underbar{1}~,
\label{DysonSurface}
\end{align}
where 
\begin{align}
\underline{\cal G}_{\,0}^{-1} =\underline{{\cal G}}_2^{-1}\circ\, ^{(-)}\underline{{\cal G}}_2^{\,\prime}
\,-\,\underline{{\cal G}}_1^{-1}\circ\, ^{(+)}\underline{{\cal G}}_1^{\,\prime}
=\frac{i}{2}\begin{pmatrix}
                                v_{1z}+v_{2z} & 2 v_{1z} h_1^+ + v_{2z} h_2^-\\
                                0 & -(v_{1z}+v_{2z})
                        \end{pmatrix} 
\label{Gsurf0}
\end{align}
and the arguments of the functions are suppressed for clarity. The surface projection $\underline{{\cal G}}$
follows from~\eqref{DysonSurface} by inverting the right hand side. In case $V_\mathrm{s}$ represents just 
an interface potential barrier, for instance, due to an epitaxial oxide layer, we are done after inversion.
For a random $V_\mathrm{s}$, however, mimicking interface roughness or scattering on surface defects,
ensemble averaging is necessary in addition to the inversion. It is the latter case to 
which we now turn in the next section.

\subsection{Derivation of the matching conditions}

Now everything is in place for deriving the matching conditions for the electron distribution functions. 
To be specific, we consider a static random surface potential $V_s$ with vanishing ensemble average, 
\begin{align}
\langle V_s(\vec{R}_\parallel)\rangle=0~, 
\end{align}
and variance 
\begin{align}
\langle V_s(\vec{R}_\parallel) V_s(\vec{R}^{\,\prime}_\parallel)\rangle
=W(\vec{R}_\parallel-\vec{R}_\parallel^{\,\prime})
&=\int\frac{d^2q_\parallel}{(2\pi)^2} 
\exp[i\vec{q}_\parallel\cdot(\vec{R}_\parallel-\vec{R}^{\,\prime}_\parallel)]\,
W(\vec{q}_\parallel)~,
\end{align}
where $W(\vec{q}_\parallel)$ is a Gaussian, whose width and height are adjustable parameters. Very often 
interfaces are not well characterized microscopically because of lack of surface diagnostics. A stochastic 
description is then mandatory. The fit parameters can then be used to reproduce experimental data
quantitatively.

Assuming $V_s$ to be weak, we iterate the inverse of~\eqref{DysonSurface} up to second order in $V_s$, 
leading, after ensemble averaging, to 
\begin{align}
\langle \underline{{\cal G}}\rangle(\vec{p}_\parallel) = 
\underline{{\cal G}}_{\,0}(\vec{p}_\parallel) + \underline{{\cal G}}_d(\vec{p}_\parallel) 
\end{align}
with 
\begin{align}
\underline{{\cal G}}_d(\vec{p}_\parallel)=
\underline{{\cal G}}_{\,0}(\vec{p}_\parallel)\,
\int\frac{d^2 q_\parallel}{(2\pi)^2} W(\vec{p}_\parallel-\vec{q}_\parallel)
\underline{{\cal G}}_{\,0}(\vec{q}_\parallel)\,
\underline{{\cal G}}_{\,0}(\vec{p}_\parallel)~.
\end{align}

In first-order gradient expansion the ensemble average of the interface 
T-matrix~\eqref{ProjectionTmatrix} becomes thus, after inserting the inverse of~\eqref{Gsurf0},
\begin{align}
\langle \underline{\tau}_{\,k} \rangle(\vec{p}_\parallel) =
\tau_{\,k}^{(b)}(\vec{p}_\parallel) + \tau_{\,k}^{(d)}(\vec{p}_\parallel)
\end{align}
with 
\begin{align}
\tau_{\,k}^{(b)}(\vec{p}_\parallel) &= \frac{2i\,v_{kz}^2}{v_{12}^2}\begin{pmatrix}
                                v_{12} & 2 s_k v_{12} - h_{12}\\
                                0 & -v_{12}
                        \end{pmatrix}
- i v_{kz} \begin{pmatrix}
             1 & s_k\\
             0 & -1
           \end{pmatrix}~,
\\
\tau_{\,k}^{(d)}(\vec{p}_\parallel)&= -\frac{8i v_{kz}^2}{v_{12}^4} \int\frac{d^2 q_\parallel}{(2\pi)^2}
W(\vec{p}_\parallel-\vec{q}_\parallel)
\bigg[\frac{v_{12}}{\bar{v}_{12}} \begin{pmatrix}
                  v_{12} & 2v_{12}s_k-2\, h_{12} \\
                  0 &  -v_{12}
                  \end{pmatrix}
+ \frac{v_{12}^2}{\bar{v}_{12}^2} \begin{pmatrix}
                  0 & \bar{h}_{12} \\
                  0 & 0
                  \end{pmatrix}
\bigg]~
\label{tau_k_averaged}
\end{align}
denoting, respectively, the ballistic ($b$) and diffusive ($d$) parts of the interface T-matrix, 
that is, the part conserving the lateral momentum and the part which does not. 
For clarity we introduced the abbreviations 
\begin{align}
v_{12} &= v_{1z}(\vec{p}_\parallel)+v_{2z}(\vec{p}_\parallel)~,~~
\bar{v}_{12} = v_{1z}(\vec{q}_\parallel)+v_{2z}(\vec{q}_\parallel)~, 
\label{v12}\\
h_{12} &= 2v_{1z}(\vec{p}_\parallel)\,h_1^+(\vec{p}_\parallel) 
+ 2v_{2z}(\vec{p}_\parallel)\,h_2^-(\vec{p}_\parallel)~,~~
\bar{h}_{12} = 2v_{1z}(\vec{q}_\parallel)\,h_1^+(\vec{q}_\parallel) 
+ 2v_{2z}(\vec{q}_\parallel)\,h_2^-(\vec{q}_\parallel)~.
\label{h12}
\end{align}
Inserting the components of $\langle \underline{\tau}_{\,k}\rangle$ into~\eqref{Inhomogeneity}, the 
inhomogeneity $U_k$ of the Boltzmann equation follows straight after some matrix algebra.

In order to obtain the Boltzmann equation in its canonical form, we follow Rammer and Smith~\cite{RS86}
and set $h_k=1-2f_k$ and $h^{(k)}=1-2f^{(k)}$, where $f_k$ and $f^{(k)}$ are the electron distribution 
functions. Introducing moreover distribution functions separately for left- and right-moving electrons, 
$f^{(k)}_\pm$ and $f_k^\pm$, and rewriting the drift term on the left hand side of~\eqref{BE1} in terms 
of $Z, T, \vec{p}_\parallel$, and $E$, the Boltzmann equation becomes
\begin{align}
\bigg[\frac{\partial}{\partial T}  \pm v_{kz} \frac{\partial}{\partial Z}\bigg]\, f^{(k)}_\pm - C^{(k)}_\pm 
= \delta(Z)W_\pm^{(k)},
\label{BE2}
\end{align}
where $C_\pm^{(k)}$ follows from~\eqref{CollIntegral} by rewriting the collision integral in terms 
of forward and backward scattering processes and distribution functions, and  
\begin{align}
W_\pm^{(k)} &= \frac{v_{kz}^2}{v_{12}^2}\,[2\tilde{h}_{12}+ 4v_{12}f_k^\pm-4 v_{12} \tilde{s}_k]  
+v_{kz}\,[\tilde{s}_k - 2 f_k^\pm] \nonumber\\
&-\frac{4\,v_{kz}^2}{v_{12}^3}\int\frac{d^2q_\parallel}{(2\pi)^2}
\frac{W(\vec{p}_\parallel-\vec{q}_\parallel)}{\bar{v}_{12}}[4\tilde{h}_{12}-4v_{12}\tilde{s}_k + 4v_{12}f^\pm_k]
\nonumber\\
&+\frac{8\,v_{kz}^2}{v_{12}^2}\int\frac{d^2q_\parallel}{(2\pi)^2}
\frac{W(\vec{p}_\parallel-\vec{q}_\parallel)}{\bar{v}_{12}^2}\tilde{\bar{h}}_{12}~ 
\label{Wk}
\end{align}
with $\tilde{s}_k$ and $\tilde{h}_{12}$ defined similarly to~\eqref{sk} and~\eqref{h12},  
but with $h^\pm_i$ replaced by $f^\pm_i$. In the last term $\tilde{h}_{12}$ depends on 
$\vec{q}_\parallel$ as indicated by the notation $\tilde{\bar{h}}_{12}$.

In applications it is better to replace the inhomogeneity by a matching/boundary condition
at $Z=0$. Hence, one solves~\eqref{BE2} for $k=1$ and $k=2$ with the right hand side set 
to zero, but with conditions connecting at $Z=0$ the solutions of the two halfspaces. 
To find the matching conditions, which are jump conditions for the distribution function 
$f^{(k)}_\pm$, we integrate~\eqref{BE2} from $Z=0^-$ to $Z=0^+$. Taking into account that, quite 
generally~\cite{Cercignani88}, the boundary condition for a Boltzmann equation can only specify 
the distribution function of inward bound particles, the physically relevant conditions are 
(suppressing all variables except the coordinate $Z$)
\begin{align}
f_{-}^{(1)}(Z=0^-) - f_-^{(1)}(Z=0^+) &= \frac{W^{(1)}_-}{v_{1z}}\bigg|_{Z=0}~,~~
\label{Jump1}\\
f_{+}^{(2)}(Z=0^+) - f_+^{(2)}(Z=0^-) &= \frac{W^{(2)}_+}{v_{2z}}\bigg|_{Z=0}~,
\label{Jump2}
\end{align}
whereas the conditions for $f_+^{(1)}$ and $f_-^{(2)}$ are unphysical and hence abandoned.

\begin{figure}[t]
\includegraphics[width=\linewidth]{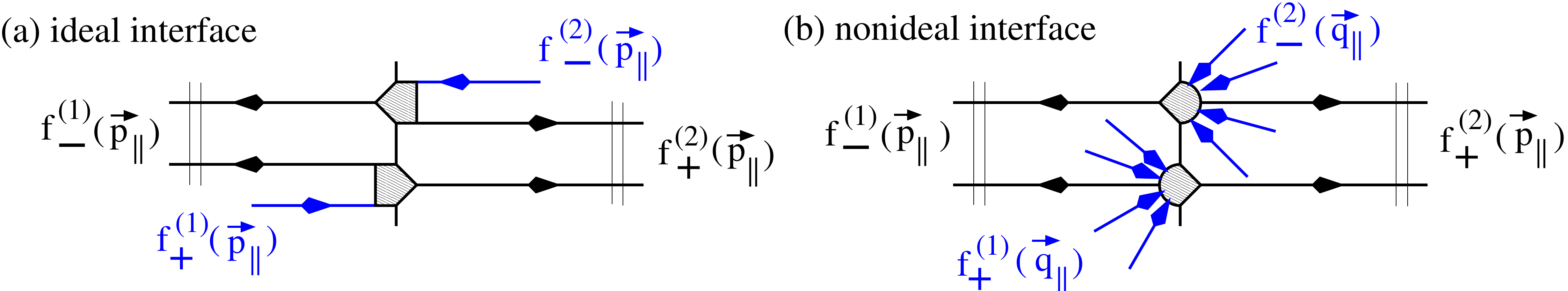}
\caption{\label{MatchingConditions}
Illustration of the matching conditions~\eqref{MC1}--\eqref{MC2} for the electron distribution 
functions at an ideal (a) and a nonideal (b) interface. The momenta $\vec{q}_\parallel$ and 
$\vec{p}_\parallel$ denote two-dimensional 
lateral momenta and the subscripts $+$ and $-$ indicate distribution functions for right- 
and left-moving electrons, that is, electrons having, respectively, momenta with positive and 
negative z-component. For the ideal interface, the lateral momentum is conserved and all the 
distribution functions involved carry the lateral momentum $\vec{p}_\parallel$. For the nonideal 
interface this is not the case and electrons with all possible lateral momenta $\vec{q}_\parallel$ 
get transmitted or reflected into a state with lateral momentum $\vec{p}_\parallel$.}  
\end{figure}

Inserting~\eqref{Wk} for $k=1$ and $k=2$ into~\eqref{Jump1} and~\eqref{Jump2}, respectively, and
taking advantage of the flexibility of extending $\underbar{G}_k$ to the unphysical side, 
which should enable us to arrange things such that (i) on the physical side of the interface 
$f_k$ is identical to $f^{(k)}$ and that (ii) at the interface $f_1^-(0)=f_-^{(1)}(0^+)$ and 
$f_2^+(0)=f_+^{(2)}(0^-)$ hold, we obtain (suppressing this time all variables except the lateral 
momentum) 
\begin{align}
f_-^{(1)}(\vec{p}_\parallel) &= [D(\vec{p}_\parallel)-\Delta D_{12}(\vec{p}_\parallel)]\, f_-^{(2)}(\vec{p}_\parallel) 
+ \int\frac{d^2q_\parallel}{(2\pi)^2}
K_{12}(\vec{p}_\parallel,\vec{q}_\parallel) f_-^{(2)}(\vec{q}_\parallel)
\nonumber\\
&+ [R(\vec{p}_\parallel) - \Delta R_{12}(\vec{p}_\parallel)]\, f_+^{(1)}(\vec{p}_\parallel) 
+ \int\frac{d^2q_\parallel}{(2\pi)^2}
K_{11}(\vec{p}_\parallel,\vec{q}_\parallel) f_+^{(1)}(\vec{q}_\parallel)~,
\label{MC1}\\
f_+^{(2)}(\vec{p}_\parallel) &=  [D(\vec{p}_\parallel) -\Delta D_{21}(\vec{p}_\parallel)]\, f_+^{(1)}(\vec{p}_\parallel) 
+ \int\frac{d^2q_\parallel}{(2\pi)^2}
K_{21}(\vec{p}_\parallel,\vec{q}_\parallel) f_+^{(1)}(\vec{q}_\parallel)
\nonumber\\
&+ [R((\vec{p}_\parallel) - \Delta R_{21}(\vec{p}_\parallel)]\, f_-^{(2)}(\vec{p}_\parallel) 
+ \int\frac{d^2q_\parallel}{(2\pi)^2}
K_{22}(\vec{p}_\parallel,\vec{q}_\parallel) f_-^{(2)}(\vec{q}_\parallel)
\label{MC2}
\end{align}
with the abbreviations 
\begin{align}
D(\vec{p}_\parallel)&=\frac{4v_{1z}v_{2z}}{v_{12}^2} = 1-R(\vec{p}_\parallel)~,
\\
\Delta D_{ij}(\vec{p}_\parallel) &=
\frac{32\,v_{iz}\,v_{jz}}{v_{12}^3} \int\frac{d^2q_\parallel}{(2\pi)^2}
\frac{W(\vec{p}_\parallel-\vec{q}_\parallel)}{\bar{v}_{12}}~,
\\
\Delta R_{ij}(\vec{p}_\parallel) &= 
\frac{16\,v_{iz}\,(v_{iz}-v_{jz})}{v_{12}^3} \int\frac{d^2q_\parallel}{(2\pi)^2}
\frac{ W(\vec{p}_\parallel-\vec{q}_\parallel)}{\bar{v}_{12}}~,
\\
K_{ij}(\vec{p}_\parallel,\vec{q}_\parallel)&=\frac{16\,v_{iz}\,\bar{v}_{jz}}{v_{12}^2}\,
\frac{W(\vec{p}_\parallel-\vec{q}_\parallel)}{\bar{v}^2_{12}}~.
\end{align}
The velocity sums $v_{12}$ and $\bar{v}_{12}$ have been defined in~\eqref{v12} 
and $\bar{v}_{iz}=v_{iz}(\vec{q}_\parallel)$. 

Equations~\eqref{MC1} and~\eqref{MC2}, specifying at the interface for each halfspace the 
distribution functions for the inward bound electrons, are our main result. From the 
structure of these equations it is clear that the distribution functions of the two 
halfspaces are coupled. Hence, a numerical solution of the interface problem requires an 
iterative approach. In each step, one has to solve for each halfspace the homogeneous part 
of~\eqref{BE2}, using~\eqref{MC1} and~\eqref{MC2} with distribution functions of the 
previous step as boundary conditions, and iterate until selfconsistency.

The terms in~\eqref{MC1} and~\eqref{MC2} involving the quantum-mechanical transmission and 
reflection probabilities, $D$ and $R$, respectively, conserve the lateral momentum. They 
are the only ones left for an ideal interface. The remaining terms are due to the 
randomness of the interface potential $V_{s}$, modeling either surface roughness or 
the scattering on surface defects. These terms do not conserve the lateral momentum 
and should thus be considered as diffusive matching/boundary conditions. An 
illustration of~\eqref{MC1} and~\eqref{MC2} is given in Fig.~\ref{MatchingConditions}. 

\section{Concluding remarks}

Starting from the atomic scale Dyson equation for the Keldysh matrix propagator, we 
derived for a planar, nonideal interface Boltzmann equations for the halfspace's
electron distribution functions together with matching conditions at the interface. 
The approach, developed by Kieselmann~\cite{Kieselmann85}, who applied it to 
superconducting proximity contacts in thermal equilibrium, is based on a pseudization 
of the interface transport problem by introducing two auxiliary, interface-free systems, 
for which the canonical procedure of obtaining Boltzmann equations from quantum kinetics 
can be applied, and a matching procedure for Green functions at the interface. Due to the 
flexibility of defining the auxiliary systems on the unphysical side of the interface, 
the distribution functions belonging to the auxiliary propagators can be eliminated. 
At the end, one is thus left with equations containing only distribution functions of 
the interface-carrying system.

The matching conditions~\eqref{MC1} and~\eqref{MC2} are not new. They have been obtained 
before. Specifically for white noise, where $W(\vec{q}_\parallel)= \mathrm{const}$, 
Eqs.~\eqref{MC1} and \eqref{MC2} are identical to the matching conditions 
derived by Dugaev and coworkers~\cite{DLP95} using Falkovsky's approach~\cite{Falkovsky83}
based on the single electron Schr\"odinger equation, an expansion of its solution
for fixed energy in terms of plane waves, and an 
identification of the distribution function with the diagonal element of the 
density matrix in momentum space. For $W(\vec{q})=0$, Eqs.~\eqref{MC1} and~\eqref{MC2}
reduce to the matching conditions of an ideal interface, containing only the 
quantum-mechanical transmission and reflection probabilities due to the potential step 
and the mismatch of the effective masses. They could be guessed in fact from flux 
conservation at the interface~\cite{Schroeder92}. Still, it is gratifying to be able 
to construct the conditions also systematically from quantum kinetics.

Pseudizing the quantum-kinetic interface transport problem is a quite powerful 
technique. From Kieselmann's work~\cite{Kieselmann85} it is clear that by extending 
the Keldysh matrix structure to Nambu space, nonideal superconducting interfaces 
in nonequilibrium can be described along the lines presented above as an alternative 
to the widespread use of transfer Hamiltonians and scattering 
matrices~\cite{CF01,GD07,ECB15}. One would then 
obtain Boltzmann-type equations and matching conditions for the $\xi-$integrated
propagators instead of the electron distribution functions. For this object the 
derivation would be moreover rather rigorous due to the presence of a Fermi scale
making the full spectral function~\eqref{SpectralFunction}, containing the real 
and imaginary parts of the selfenergy, a peaked function in 
$\xi=\vec{p}^{\,2}/2m_*$ in case energy is measured from the Fermi energy. Even for 
semiconductors some of the interaction effects buried in~\eqref{SpectralFunction}, 
leading to a renormalization of the electronic quasiparticle, can be most probably 
treated within the extended quasiparticle approximation~\cite{SL94,SL95}. As long 
as the selfenergy corrections are such that a quasiparticle dispersion can still
be sufficiently well defined, the derivation of the matching conditions goes
through smoothly. It depends however on the details of the interaction 
process. A case by case study is thus required. It is also conceivable to treat the 
randomness of the interface potential beyond second order perturbation theory. 
Equation~\eqref{DysonSurface} has the structure of a Dyson equation for noninteracting 
electrons in a random potential. Using $\underline{{\cal G}}_{\,0}$ defined 
in~\eqref{Gsurf0} as the unperturbed, bare propagator, a diagrammatic approach can 
thus be envisaged to treat arbitrarily strong interface roughness or strong scattering 
on surface defects. Finally, it is straightforward to augment the matrix structure 
by the electron spin, enabling thereby the construction of matching conditions for 
magnetically active interfaces.

\vspace{3mm}
Support from the Deutsche Forschungsgemeinschaft through 
project 495729137 is greatly acknowledged. 


\bibliography{./ref}

\begin{thebibliography}{38}
\expandafter\ifx\csname natexlab\endcsname\relax\def\natexlab#1{#1}\fi
\expandafter\ifx\csname bibnamefont\endcsname\relax
  \def\bibnamefont#1{#1}\fi
\expandafter\ifx\csname bibfnamefont\endcsname\relax
  \def\bibfnamefont#1{#1}\fi
\expandafter\ifx\csname citenamefont\endcsname\relax
  \def\citenamefont#1{#1}\fi
\expandafter\ifx\csname url\endcsname\relax
  \def\url#1{\texttt{#1}}\fi
\expandafter\ifx\csname urlprefix\endcsname\relax\def\urlprefix{URL }\fi
\providecommand{\bibinfo}[2]{#2}
\providecommand{\eprint}[2][]{\url{#2}}

\bibitem[{\citenamefont{Cercignani}(1988)}]{Cercignani88}
\bibinfo{author}{\bibfnamefont{C.}~\bibnamefont{Cercignani}},
  \emph{\bibinfo{title}{The Boltzmann equation and its applications}}
  (\bibinfo{publisher}{Springer-Verlag}, \bibinfo{address}{New York},
  \bibinfo{year}{1988}).

\bibitem[{\citenamefont{Kogan}(1969)}]{Kogan69}
\bibinfo{author}{\bibfnamefont{M.~N.} \bibnamefont{Kogan}},
  \emph{\bibinfo{title}{Rarefied gas dynamics}} (\bibinfo{publisher}{Plenum
  Press}, \bibinfo{address}{New York}, \bibinfo{year}{1969}).

\bibitem[{\citenamefont{Ecker}(1972)}]{Ecker72}
\bibinfo{author}{\bibfnamefont{G.}~\bibnamefont{Ecker}},
  \emph{\bibinfo{title}{Theory of fully ionized plasmas}}
  (\bibinfo{publisher}{Academic Press}, \bibinfo{address}{New York},
  \bibinfo{year}{1972}).

\bibitem[{\citenamefont{Smith and Jensen}(1989)}]{SJ89}
\bibinfo{author}{\bibfnamefont{H.}~\bibnamefont{Smith}} \bibnamefont{and}
  \bibinfo{author}{\bibfnamefont{H.~H.} \bibnamefont{Jensen}},
  \emph{\bibinfo{title}{Transport Phenomena}} (\bibinfo{publisher}{Clarendon
  Press}, \bibinfo{address}{Oxford}, \bibinfo{year}{1989}).

\bibitem[{\citenamefont{Bogolyubov}(1946)}]{Bogolyubov46}
\bibinfo{author}{\bibfnamefont{N.~N.} \bibnamefont{Bogolyubov}},
  \emph{\bibinfo{title}{The Problems of Dynamical Theory in Statistical
  Physics}} (\bibinfo{publisher}{Gostechizdat},
  \bibinfo{address}{Moscow-Leningrad}, \bibinfo{year}{1946}).

\bibitem[{\citenamefont{Bogolyubov and Gurov}(1947)}]{BG47}
\bibinfo{author}{\bibfnamefont{N.~N.} \bibnamefont{Bogolyubov}}
  \bibnamefont{and} \bibinfo{author}{\bibfnamefont{K.~P.} \bibnamefont{Gurov}},
  \bibinfo{journal}{J. Exp. Theor. Phys.} \textbf{\bibinfo{volume}{17}},
  \bibinfo{pages}{614} (\bibinfo{year}{1947}).

\bibitem[{\citenamefont{Kadanoff and Baym}(1962)}]{KB62}
\bibinfo{author}{\bibfnamefont{L.~P.} \bibnamefont{Kadanoff}} \bibnamefont{and}
  \bibinfo{author}{\bibfnamefont{G.}~\bibnamefont{Baym}},
  \emph{\bibinfo{title}{Quantum statistical mechanics}}
  (\bibinfo{publisher}{Addison-Wesley}, \bibinfo{address}{Redwood City},
  \bibinfo{year}{1962}).

\bibitem[{\citenamefont{Keldysh}(1965)}]{Keldysh65}
\bibinfo{author}{\bibfnamefont{L.~V.} \bibnamefont{Keldysh}},
  \bibinfo{journal}{Sov. Phys. JETP} \textbf{\bibinfo{volume}{20}},
  \bibinfo{pages}{1018} (\bibinfo{year}{1965}).

\bibitem[{\citenamefont{Bonitz}(2016)}]{Bonitz2016}
\bibinfo{author}{\bibfnamefont{M.}~\bibnamefont{Bonitz}},
  \emph{\bibinfo{title}{Quantum kinetic theory}}
  (\bibinfo{publisher}{Springer}, \bibinfo{address}{Cham},
  \bibinfo{year}{2016}).

\bibitem[{\citenamefont{Rammer and Smith}(1986)}]{RS86}
\bibinfo{author}{\bibfnamefont{J.}~\bibnamefont{Rammer}} \bibnamefont{and}
  \bibinfo{author}{\bibfnamefont{H.}~\bibnamefont{Smith}},
  \bibinfo{journal}{Rev. Mod. Phys.} \textbf{\bibinfo{volume}{58}},
  \bibinfo{pages}{323} (\bibinfo{year}{1986}).

\bibitem[{\citenamefont{Landau}(1958)}]{Landau58}
\bibinfo{author}{\bibfnamefont{L.~D.} \bibnamefont{Landau}},
  \bibinfo{journal}{Sov. Phys. JETP} \textbf{\bibinfo{volume}{3}},
  \bibinfo{pages}{920} (\bibinfo{year}{1958}).

\bibitem[{\citenamefont{Prange and Kadanoff}(1964)}]{PK64}
\bibinfo{author}{\bibfnamefont{R.~E.} \bibnamefont{Prange}} \bibnamefont{and}
  \bibinfo{author}{\bibfnamefont{L.~P.} \bibnamefont{Kadanoff}},
  \bibinfo{journal}{Phys. Rev.} \textbf{\bibinfo{volume}{134}},
  \bibinfo{pages}{A566} (\bibinfo{year}{1964}).

\bibitem[{\citenamefont{Serene and Rainer}(1983)}]{SR83}
\bibinfo{author}{\bibfnamefont{J.~W.} \bibnamefont{Serene}} \bibnamefont{and}
  \bibinfo{author}{\bibfnamefont{D.}~\bibnamefont{Rainer}},
  \bibinfo{journal}{Phys. Rep.} \textbf{\bibinfo{volume}{101}},
  \bibinfo{pages}{221} (\bibinfo{year}{1983}).

\bibitem[{\citenamefont{\v{S}pi\v{c}ka and Lipavsk\'y}(1994)}]{SL94}
\bibinfo{author}{\bibfnamefont{V.}~\bibnamefont{\v{S}pi\v{c}ka}}
  \bibnamefont{and}
  \bibinfo{author}{\bibfnamefont{P.}~\bibnamefont{Lipavsk\'y}},
  \bibinfo{journal}{Phys. Rev. Lett.} \textbf{\bibinfo{volume}{73}},
  \bibinfo{pages}{3439} (\bibinfo{year}{1994}).

\bibitem[{\citenamefont{\v{S}pi\v{c}ka and Lipavsk\'y}(1995)}]{SL95}
\bibinfo{author}{\bibfnamefont{V.}~\bibnamefont{\v{S}pi\v{c}ka}}
  \bibnamefont{and}
  \bibinfo{author}{\bibfnamefont{P.}~\bibnamefont{Lipavsk\'y}},
  \bibinfo{journal}{Phys. Rev. B} \textbf{\bibinfo{volume}{52}},
  \bibinfo{pages}{14615} (\bibinfo{year}{1995}).

\bibitem[{\citenamefont{Roth}(1992)}]{Roth92}
\bibinfo{author}{\bibfnamefont{L.}~\bibnamefont{Roth}}, in
  \emph{\bibinfo{booktitle}{Handbook on Semiconductors}}, edited by
  \bibinfo{editor}{\bibfnamefont{T.~S.} \bibnamefont{Moss}}
  (\bibinfo{publisher}{Elsevier Science Publishers B. V.},
  \bibinfo{address}{Amsterdam, NL}, \bibinfo{year}{1992}), p.
  \bibinfo{pages}{489}.

\bibitem[{\citenamefont{Smirnov and Jungemann}(2006)}]{SJ06}
\bibinfo{author}{\bibfnamefont{S.}~\bibnamefont{Smirnov}} \bibnamefont{and}
  \bibinfo{author}{\bibfnamefont{C.}~\bibnamefont{Jungemann}},
  \bibinfo{journal}{J. Appl. Phys.} \textbf{\bibinfo{volume}{99}},
  \bibinfo{pages}{063707} (\bibinfo{year}{2006}).

\bibitem[{\citenamefont{Fischetti and Lax}(1988)}]{FL88}
\bibinfo{author}{\bibfnamefont{M.~V.} \bibnamefont{Fischetti}}
  \bibnamefont{and} \bibinfo{author}{\bibfnamefont{S.~E.} \bibnamefont{Lax}},
  \bibinfo{journal}{Phys. Rev. B} \textbf{\bibinfo{volume}{38}},
  \bibinfo{pages}{9721} (\bibinfo{year}{1988}).

\bibitem[{\citenamefont{Rupp et~al.}(2016)\citenamefont{Rupp, Jungemann, Hong,
  Bina, Grasser, and J\"ungel}}]{RJH16}
\bibinfo{author}{\bibfnamefont{K.}~\bibnamefont{Rupp}},
  \bibinfo{author}{\bibfnamefont{C.}~\bibnamefont{Jungemann}},
  \bibinfo{author}{\bibfnamefont{S.-M.} \bibnamefont{Hong}},
  \bibinfo{author}{\bibfnamefont{M.}~\bibnamefont{Bina}},
  \bibinfo{author}{\bibfnamefont{T.}~\bibnamefont{Grasser}}, \bibnamefont{and}
  \bibinfo{author}{\bibfnamefont{A.}~\bibnamefont{J\"ungel}},
  \bibinfo{journal}{J. Comput. Electron.} \textbf{\bibinfo{volume}{15}},
  \bibinfo{pages}{939} (\bibinfo{year}{2016}).

\bibitem[{\citenamefont{Zaitsev}(1984)}]{Zaitsev84}
\bibinfo{author}{\bibfnamefont{A.~V.} \bibnamefont{Zaitsev}},
  \bibinfo{journal}{Sov. Phys. JETP} \textbf{\bibinfo{volume}{59}},
  \bibinfo{pages}{1015} (\bibinfo{year}{1984}).

\bibitem[{\citenamefont{Kieselmann}(1985)}]{Kieselmann85}
\bibinfo{author}{\bibfnamefont{G.}~\bibnamefont{Kieselmann}},
  \emph{\bibinfo{title}{Quasiklassische Theorie supraleitender
  Metall-Metall-Kontakte}} (\bibinfo{publisher}{PhD thesis},
  \bibinfo{address}{Universit\"at Bayreuth}, \bibinfo{year}{1985}).

\bibitem[{\citenamefont{Millis et~al.}(1988)\citenamefont{Millis, Rainer, and
  Sauls}}]{MRS88}
\bibinfo{author}{\bibfnamefont{A.}~\bibnamefont{Millis}},
  \bibinfo{author}{\bibfnamefont{D.}~\bibnamefont{Rainer}}, \bibnamefont{and}
  \bibinfo{author}{\bibfnamefont{J.~A.} \bibnamefont{Sauls}},
  \bibinfo{journal}{Phys. Rev. B} \textbf{\bibinfo{volume}{38}},
  \bibinfo{pages}{4504} (\bibinfo{year}{1988}).

\bibitem[{\citenamefont{Ashauer et~al.}(1986)\citenamefont{Ashauer, Kieselmann,
  and Rainer}}]{AKR86}
\bibinfo{author}{\bibfnamefont{B.}~\bibnamefont{Ashauer}},
  \bibinfo{author}{\bibfnamefont{G.}~\bibnamefont{Kieselmann}},
  \bibnamefont{and} \bibinfo{author}{\bibfnamefont{D.}~\bibnamefont{Rainer}},
  \bibinfo{journal}{J. Low Temp. Phys.} \textbf{\bibinfo{volume}{63}},
  \bibinfo{pages}{349} (\bibinfo{year}{1986}).

\bibitem[{\citenamefont{Kieselmann}(1987)}]{Kieselmann87}
\bibinfo{author}{\bibfnamefont{G.}~\bibnamefont{Kieselmann}},
  \bibinfo{journal}{Phys. Rev. B} \textbf{\bibinfo{volume}{35}},
  \bibinfo{pages}{6762} (\bibinfo{year}{1987}).

\bibitem[{\citenamefont{Cuevas and Fogelstr\"om}(2001)}]{CF01}
\bibinfo{author}{\bibfnamefont{J.~C.} \bibnamefont{Cuevas}} \bibnamefont{and}
  \bibinfo{author}{\bibfnamefont{M.}~\bibnamefont{Fogelstr\"om}},
  \bibinfo{journal}{Phys. Rev. B} \textbf{\bibinfo{volume}{64}},
  \bibinfo{pages}{104502} (\bibinfo{year}{2001}).

\bibitem[{\citenamefont{Graser and Dahm}(2007)}]{GD07}
\bibinfo{author}{\bibfnamefont{S.}~\bibnamefont{Graser}} \bibnamefont{and}
  \bibinfo{author}{\bibfnamefont{T.}~\bibnamefont{Dahm}},
  \bibinfo{journal}{Phys. Rev. B} \textbf{\bibinfo{volume}{75}},
  \bibinfo{pages}{014507} (\bibinfo{year}{2007}).

\bibitem[{\citenamefont{Eschrig et~al.}(2015)\citenamefont{Eschrig, Cottet,
  Belzig, and Linder}}]{ECB15}
\bibinfo{author}{\bibfnamefont{M.}~\bibnamefont{Eschrig}},
  \bibinfo{author}{\bibfnamefont{A.}~\bibnamefont{Cottet}},
  \bibinfo{author}{\bibfnamefont{W.}~\bibnamefont{Belzig}}, \bibnamefont{and}
  \bibinfo{author}{\bibfnamefont{J.}~\bibnamefont{Linder}},
  \bibinfo{journal}{New J. Phys.} \textbf{\bibinfo{volume}{17}},
  \bibinfo{pages}{083037} (\bibinfo{year}{2015}).

\bibitem[{\citenamefont{Falkovsky}(1983)}]{Falkovsky83}
\bibinfo{author}{\bibfnamefont{L.~A.} \bibnamefont{Falkovsky}},
  \bibinfo{journal}{Adv. Phys.} \textbf{\bibinfo{volume}{32}},
  \bibinfo{pages}{753} (\bibinfo{year}{1983}).

\bibitem[{\citenamefont{Dugaev et~al.}(1995)\citenamefont{Dugaev, Litvinov, and
  Petrov}}]{DLP95}
\bibinfo{author}{\bibfnamefont{V.~K.} \bibnamefont{Dugaev}},
  \bibinfo{author}{\bibfnamefont{V.~I.} \bibnamefont{Litvinov}},
  \bibnamefont{and} \bibinfo{author}{\bibfnamefont{P.~P.}
  \bibnamefont{Petrov}}, \bibinfo{journal}{Phys. Rev. B}
  \textbf{\bibinfo{volume}{52}}, \bibinfo{pages}{5306} (\bibinfo{year}{1995}).

\bibitem[{\citenamefont{Schroeder}(1992)}]{Schroeder92}
\bibinfo{author}{\bibfnamefont{D.}~\bibnamefont{Schroeder}},
  \bibinfo{journal}{J. Appl. Phys.} \textbf{\bibinfo{volume}{72}},
  \bibinfo{pages}{964} (\bibinfo{year}{1992}).

\bibitem[{\citenamefont{Bronold and Fehske}(2017)}]{BF17}
\bibinfo{author}{\bibfnamefont{F.~X.} \bibnamefont{Bronold}} \bibnamefont{and}
  \bibinfo{author}{\bibfnamefont{H.}~\bibnamefont{Fehske}},
  \bibinfo{journal}{J. Phys. D: Appl. Phys.} \textbf{\bibinfo{volume}{50}},
  \bibinfo{pages}{294003} (\bibinfo{year}{2017}).

\bibitem[{\citenamefont{Eden et~al.}(2013)\citenamefont{Eden, Park, Cho, Kim,
  Houlahan, Li, Kim, Kim, Lee, Kim et~al.}}]{EPC13}
\bibinfo{author}{\bibfnamefont{J.~G.} \bibnamefont{Eden}},
  \bibinfo{author}{\bibfnamefont{S.-J.} \bibnamefont{Park}},
  \bibinfo{author}{\bibfnamefont{J.~H.} \bibnamefont{Cho}},
  \bibinfo{author}{\bibfnamefont{M.~H.} \bibnamefont{Kim}},
  \bibinfo{author}{\bibfnamefont{T.~J.} \bibnamefont{Houlahan}},
  \bibinfo{author}{\bibfnamefont{B.}~\bibnamefont{Li}},
  \bibinfo{author}{\bibfnamefont{E.~S.} \bibnamefont{Kim}},
  \bibinfo{author}{\bibfnamefont{T.~L.} \bibnamefont{Kim}},
  \bibinfo{author}{\bibfnamefont{S.~K.} \bibnamefont{Lee}},
  \bibinfo{author}{\bibfnamefont{K.~S.} \bibnamefont{Kim}},
  \bibnamefont{et~al.}, \bibinfo{journal}{IEEE Trans. Plasma Sci.}
  \textbf{\bibinfo{volume}{41}}, \bibinfo{pages}{661} (\bibinfo{year}{2013}).

\bibitem[{\citenamefont{Garc\'ia-Moliner and Rubio}(1971)}]{GR71}
\bibinfo{author}{\bibfnamefont{F.}~\bibnamefont{Garc\'ia-Moliner}}
  \bibnamefont{and} \bibinfo{author}{\bibfnamefont{J.}~\bibnamefont{Rubio}},
  \bibinfo{journal}{Proc. Roy. Soc. Lond. A.} \textbf{\bibinfo{volume}{324}},
  \bibinfo{pages}{257} (\bibinfo{year}{1971}).

\bibitem[{\citenamefont{Eilenberger}(1968)}]{Eilenberger68}
\bibinfo{author}{\bibfnamefont{G.}~\bibnamefont{Eilenberger}},
  \bibinfo{journal}{Zeitschrift f. Physik} \textbf{\bibinfo{volume}{214}},
  \bibinfo{pages}{195} (\bibinfo{year}{1968}).

\bibitem[{\citenamefont{Zhang and Levy}(1998)}]{ZL98}
\bibinfo{author}{\bibfnamefont{S.}~\bibnamefont{Zhang}} \bibnamefont{and}
  \bibinfo{author}{\bibfnamefont{P.~M.} \bibnamefont{Levy}},
  \bibinfo{journal}{Phys. Rev. B} \textbf{\bibinfo{volume}{57}},
  \bibinfo{pages}{5336} (\bibinfo{year}{1998}).

\bibitem[{\citenamefont{Amin and Stiles}(2016{\natexlab{a}})}]{AS16a}
\bibinfo{author}{\bibfnamefont{V.~P.} \bibnamefont{Amin}} \bibnamefont{and}
  \bibinfo{author}{\bibfnamefont{M.~D.} \bibnamefont{Stiles}},
  \bibinfo{journal}{Phys. Rev. B} \textbf{\bibinfo{volume}{94}},
  \bibinfo{pages}{104419} (\bibinfo{year}{2016}{\natexlab{a}}).

\bibitem[{\citenamefont{Amin and Stiles}(2016{\natexlab{b}})}]{AS16b}
\bibinfo{author}{\bibfnamefont{V.~P.} \bibnamefont{Amin}} \bibnamefont{and}
  \bibinfo{author}{\bibfnamefont{M.~D.} \bibnamefont{Stiles}},
  \bibinfo{journal}{Phys. Rev. B} \textbf{\bibinfo{volume}{94}},
  \bibinfo{pages}{104420} (\bibinfo{year}{2016}{\natexlab{b}}).

\bibitem[{\citenamefont{Shokri and Khiabanian}(2011)}]{SK11}
\bibinfo{author}{\bibfnamefont{A.~A.} \bibnamefont{Shokri}} \bibnamefont{and}
  \bibinfo{author}{\bibfnamefont{T.}~\bibnamefont{Khiabanian}},
  \bibinfo{journal}{Int. J. Thermal Sci.} \textbf{\bibinfo{volume}{50}},
  \bibinfo{pages}{1900} (\bibinfo{year}{2011}).

\end{thebibliography}
\bibliographystyle{apsrev}

\end{document}